\documentclass[manuscript,screen, nonacm]{acmart}
\AtBeginDocument{%
  \providecommand\BibTeX{{%
    \normalfont B\kern-0.5em{\scshape i\kern-0.25em b}\kern-0.8em\TeX}}}

\setcopyright{acmlicensed}
\copyrightyear{2018}
\acmYear{2018}
\acmDOI{XXXXXXX.XXXXXXX}

\acmConference[Conference acronym 'XX]{Make sure to enter the correct
  conference title from your rights confirmation email}{June 03--05,
  2018}{Woodstock, NY}
\acmBooktitle{Woodstock '18: ACM Symposium on Neural Gaze Detection,
 June 03--05, 2018, Woodstock, NY} 
\acmISBN{978-1-4503-XXXX-X/18/06}

\begin{document}

\title{“And this is where we fu***d up!” Lessons learned from Participatory Design in Digital Civic Initiatives}

\author{Clara Rosa Cardoso}
\email{clara.rosacardoso@uni-siegen.de}
\orcid{0009-0001-8497-6400}

\affiliation{%
  \institution{University of Siegen}
  \city{Siegen}
  \country{Germany}
}

\author{Sarah Rüller}
\affiliation{%
  \institution{University of Siegen}
  \city{Siegen}
  \country{Germany}}
\email{sarah.rueller@uni-siegen.de}
\orcid{0000-0003-2541-2776}

\author{Ana O Henriques}
\affiliation{%
\institution{ITI/LARSyS, Instituto Superior Técnico, Universidade de Lisboa}
 \country{Portugal}}
\orcid{0000-0001-7844-2157}

\author{Anna R L Carter}
\affiliation{%
    \institution{Northumbria University}
 \country{UK}}
\orcid{0000-0002-2436-666X}

\author{Markus Rohde}
\affiliation{%
  \institution{University of Siegen}
  \city{Siegen}
  \country{Germany}
}

\renewcommand{\shortauthors}{Rosa Cardoso et al.}

\begin{abstract}
 Participatory design in digital civics aims to foster mutual learning and co-creation between public services and citizens. However, rarely do we collectively explore the challenges and failures we experience within PD and digital civics, to enable us to grow as a community. This workshop will explore real-world experiences that had to adapt to unforeseen circumstances. Through case presentations and thematic group discussions, participants will reflect on the challenges faced, the causes that led to these challenges, and collaboratively problem-solve effective solutions. Furthermore, we aim to discuss well-being impact on researchers and communities when faced with these obstacles, the strategies participants use to overcome them and how this can be fed back into the digital civics community. By that, the workshop seeks to foster dialogue, reflection, and collective learning, empowering participants with insights to navigate complexities effectively and promote resilient design practices in digital civics.
\end{abstract}



\received{19 April 2024}
\received[revised]{12 March 2009}
\received[accepted]{5 June 2009}
\maketitle
\section{Introduction}
Participatory design (PD) for digital civics focuses on fostering relationships between public services/officials and citizens within participatory experiences built on mutual learning, empowerment, and co-creation \cite{olivier2015digital}. The use of PD in Human-Computer Interaction (HCI) has been instrumental in articulating civic concerns and engaging the public in addressing them \cite{dow2018between}. Scholarship has underscored the challenges and potential of PD, with studies analysing various aspects such as the value of participation in community-based technology co-design \cite{reynolds2020community} and the transformative impact of digital media on civic expression and action among youth \cite{mirra2017civic}. Additionally, research has highlighted the role of digital participatory games in engaging citizens in civic activities like urban planning \cite{poplin2014digital}, the effectiveness of platforms in facilitating participatory budgeting \cite{menendez2022designing}, and implications of the digital age in public participation and in building digital social capital \cite{mandarano2010building}. However, collaborative design processes in the field often face unforeseen challenges and failure. Focusing on failure and discontinuation in digital civics in HCI, Hamm et al. \cite{hamm2024does} advocated for reframing failure as a productive learning opportunity. They proposed exploring deeper underlying factors beyond apparent obstacles like technical issues or resource scarcity, and re-conceptualising civic tech as sociotechnical infrastructuring. Exploring PD within a community-based mobile learning initiative, Richardson \cite{richardson2023failed} reflected on the successes and challenges of digital civics, emphasising the need to avoid inadvertently supporting austerity measures and shifting responsibilities onto individuals. 

Despite the established role of participatory approaches in shaping digital technologies for societal benefit, the ongoing challenge to understand and navigate the complexities involved in their implementation remains. This workshop emerges from recognising the inherent and still under-discussed occurrence of unexpected outcomes and failure within collaborative design processes in digital civics. While previous workshops have also explored relevant issues such as the role and practices of PD facilitators \cite{dahl2022towards} and the temporal aspects of the design process \cite{nouwens2022time}, there is still a notable gap in the literature regarding the exploration of failure within PD and digital civics. Acknowledging that such initiatives often encounter unforeseen obstacles and diverge from initial plans, this workshop is motivated by the imperative to address these difficulties effectively. By examining real-world cases and exploring the potential impact of unexpected outcomes in digital civics initiatives, we seek to help illuminate the intricacies of collaborative design dynamics. Through our central question: \textit{Can failure in PD for digital civics be a catalyst for positive change, and if so, how can we best leverage it?}, the workshop aims to bridge distances and diversities by stimulating dialogue and collective reflection on the role of failure in driving innovation in PD and learning within digital civics. Ultimately, we hope to empower participants with insights and strategies for tackling these complexities, fostering a culture of resilient and impactful design practices.

\section{WORKSHOP DESCRIPTION AND STRUCTURE}
This half-day physical workshop is open for 10 to 20 participants. A detailed description of planned activities can be found below. By exploring real, hyperlocal cases, we aim to create a space for exchanging experiences and sparking connections within the participants. Working and discussing in smaller groups to dissect the specific themes that emerged, should foster connections and building a shared understanding through collaborative exploration. Key discussion points will be documented on whiteboards, setting the stage for the following plenary discussion where we can collectively analyse the findings.
Finally, groups will reconvene to share their main ideas in a plenary session, addressing questions like: {\itshape What can we learn from failure? What kind of knowledge do these experiences yield? How do we efficiently respond to and recover from setbacks?} The overarching goal of the workshop is to foster a deeper understanding of collaborative design complexities in digital civics and identify strategies for more effective practices. After the workshop, participants can continue the conversation using the whiteboard on the workshop website, and the already existing DCitizens Discord server and Seminar Group (our ongoing \href{https://dcitizens.eu/}{EU-funded project}). On-Site requirements for the workshop execution includes a room for up to 25 people, seating and tables for five groups, along with a whiteboard, projector and screen with sound system, HDMI connectors, and five flip charts with pens for group activities.

\subsection{Planned Activities}
\begin{itemize}
\item {\itshape Introduction \& welcome} (15 minutes) - Introduction to the workshop and outline of agenda and goals.
\item {\itshape Round of introductions} (up to 20 minutes) - Each participant will have one minute to present their research domain and interest in the workshop topic.
\item {\itshape Presentation of real-world cases} (up to 60 minutes) - Three to four presenters will showcase their projects, focusing on lessons learned, especially setbacks and fallouts.
\item {\itshape Coffee break} (15 minutes)
\item {\itshape Thematic group discussions} (45 minutes) - Participants will form three to five groups, each exploring different aspects defined by consensus. Topics, based on the real-world cases presented, may include types of digressions, root causes analysis, mitigation strategies, practical implications, adaptation vs. persistence, and impact on researchers.
\item {\itshape Plenary session} (45 minutes) - All participants come together. Each group will have five minutes to share in plenum the main points discussed. Approximately three questions can be addressed following each group presentation.
\item {\itshape Shaping tomorrow} (15 minutes) - Groups will be invited to quickly add to a common list (whiteboard) one final actionable insight they recommend for future civic initiatives through PD. Each contribution can be accompanied by a one-minute explanation. 
\item {\itshape Final thoughts \& wrap-up}  (15 minutes) - End of day thoughts and takeaways.
\end{itemize}



\section{WORKSHOP PARTICIPATION}
The workshop aims to engage participants from diverse backgrounds. Given its open questions and lessons-learned nature, it is equally fit to practitioners, researchers, and scholars interested in discussing practical aspects of embedded or participatory design within digital civics.\newline
\textbf{Promotional strategy:} In early June 2024, a website will promote the workshop and offer additional information to potential participants. Building upon the foundations laid by the ongoing Horizon project \cite{dcitizensproj22} there is a network of researchers, practitioners, community members, and policymakers who share an interest in digital civics that we can reach out to. We will also leverage related workshops in which our organisers are involved \cite{Henriques2024, carter24sig, carter24pdc}, to promote our event to interested audiences. 
\newline
\textbf{Recruiting and selecting participants:} Interested parties should send one to two-page submissions, stating their contribution and motivation related to the workshop. We welcome HCI and design researchers, as well as practitioners involved in or interested in various aspects of digital civics. Submissions that include anecdotes about setbacks or failures in PD efforts are encouraged, as these may serve as case studies during the workshop and in the subsequent article. Unorthodox formats are welcome, such as (but not limited to) video pitch, pictorial or storyboard presentation. Submissions must adhere to the accessibility guidelines of NordiCHI 2024. Participants are selected based on their experience relevance and/or interest for the workshop.

\subsection{Link to the call for participation}
The call for participation can be accessed here: \url{https://dcitizens.eu/nordichi-2024/}
\section{ORGANIZERS}

\textbf{Clara Rosa Cardoso} is a PhD student and research assistant at the Institute for Information Systems and New Media at the University of Siegen. She currently contributes to research projects involving PD initiatives for digital civics with underserved communities in Portugal and computer clubs for children in refugee camps in Palestine.

\textbf{Dr. Sarah Rüller} is a PostDoc researcher at the Institute for Information Systems and New Media at the University of Siegen. Her doctoral research focuses on computer clubs and computer-supported project-based learning with Imazighen (indigenous inhabitants of Morocco) in the High Atlas. She is also focussing on censorship on social media platforms, particularly in the context Palestine/Israel.

\textbf{Ana O. Henriques} is currently a junior researcher at the Interactive Technologies Institute / LARSyS, at the University of Lisbon. Ana has focused their research at the intersections of ethics, feminist HCI and digital civics. They are developing the concept of community-led ethics as a process of feminist ethical frameworking for digital civics in the context of the DCitizens project.

\textbf{Dr. Anna Carter} is an Innovation Fellow at Northumbria University, she has extensive experience in designing technologies for local council regeneration programs, with a focus on creating accessible digital experiences using human-centred methods and participatory design. She currently contributes to the Digital Civics research capacities for early career researchers in the EU-funded DCitizens Programme and on digital civics, outdoor spaces and sense of place as part of the EPSRC funded Centre for Digital Citizens.

\textbf{(apl.) Prof. Dr. Markus Rohde}, is a founding member of the International Institute for Socio-Informatics and a co-editor of the International Reports on Socio-Informatics (IRSI). As head of Community Informatics at the Institute for Information Systems and New Media at the University of Siegen, his work focuses on Human-Computer Interaction, Computer Supported Cooperative Work (CSCW), organizational and collaborative learning, virtual teams, NGOs and (new) social movements.

\section{INTENDED OUTCOMES}
Based on the considerations outlined, the workshop has three major intended outcomes.
We want to bring together different stakeholders with distinct views to learn from mistakes in PD in digital civics and how such lessons can help the field, its practices and research move forward while building long-term, sustainable, transdisciplinary relationships. This initiative aims to \textbf{enhance collaboration and facilitate exchange of diverse perspectives within the digital civics community}. Such efforts can help equip future PDers with the tools to navigate the unexpected developments inherent in the field. This workshop wishes to \textbf{inspire collective reflection on current design and research practices within digital civics}. Participants are encouraged to engage critically with research methods and expectations, drawing insights from both successes and mistakes. With a focus on failures, we seek to identify strategies for more effective design processes. By reflecting on each other’s experiences, we aim to enhance our ability to navigate serendipity and improve our practice. We intend to \textbf{communicate the outcomes of our discussions in a collaborative paper on PD work with digital civics}, in an HCI-focused journal. Moreover, a dedicated webpage will host workshop materials, including presentations, literature and outcomes from the groups and plenum. A forum will facilitate ongoing discussions beyond the conference, engaging a broader audience in the conversation on lessons learned from misfortunes in civic initiatives through PD.

\section{Acknowledgments}
We thank our funding bodies at the European Commission (101079116 Fostering Digital Civics Research and Innovation
in Lisbon), EPSRC (EP/T022582/1 Centre for Digital Citizens - Next Stage Digital Economy Centre) and the Portuguese
Recovery and Resilience Program (PRR), IAPMEI/ANI/FCT under Agenda C645022399-00000057 (eGamesLab).




\bibliographystyle{ACM-Reference-Format}
\bibliography{references.bib}


\begin{thebibliography}{15}


\ifx \showCODEN    \undefined \def \showCODEN     #1{\unskip}     \fi
\ifx \showDOI      \undefined \def \showDOI       #1{#1}\fi
\ifx \showISBNx    \undefined \def \showISBNx     #1{\unskip}     \fi
\ifx \showISBNxiii \undefined \def \showISBNxiii  #1{\unskip}     \fi
\ifx \showISSN     \undefined \def \showISSN      #1{\unskip}     \fi
\ifx \showLCCN     \undefined \def \showLCCN      #1{\unskip}     \fi
\ifx \shownote     \undefined \def \shownote      #1{#1}          \fi
\ifx \showarticletitle \undefined \def \showarticletitle #1{#1}   \fi
\ifx \showURL      \undefined \def \showURL       {\relax}        \fi
\providecommand\bibfield[2]{#2}
\providecommand\bibinfo[2]{#2}
\providecommand\natexlab[1]{#1}
\providecommand\showeprint[2][]{arXiv:#2}

\bibitem[Carter et~al\mbox{.}(2024a)]%
        {carter24pdc}
\bibfield{author}{\bibinfo{person}{Anna R.~L. Carter}, \bibinfo{person}{Kyle Montague}, \bibinfo{person}{Reem Talhouk}, \bibinfo{person}{Ana~O. Henriques}, \bibinfo{person}{Hugo Nicolau}, \bibinfo{person}{Tiffany Knearem}, \bibinfo{person}{Ceylan Besevli}, \bibinfo{person}{Firaz Peer}, \bibinfo{person}{Clara Crivellaro}, {and} \bibinfo{person}{Sarah Rüller}.} \bibinfo{year}{2024}\natexlab{a}.
\newblock \showarticletitle{Envisioning Collaborative Futures: Advancing the Frontiers of Embedded Research.}. In \bibinfo{booktitle}{\emph{18th Biennial Participatory Design Conference}} \emph{(\bibinfo{series}{PDC ’24})}. \bibinfo{publisher}{Association for Computing Machinery}, \bibinfo{address}{New York, NY, USA}.
\newblock
\urldef\tempurl%
\url{https://doi.org/10.1145/3661455.3669890}
\showDOI{\tempurl}


\bibitem[Carter et~al\mbox{.}(2024b)]%
        {carter24sig}
\bibfield{author}{\bibinfo{person}{Anna R.~L. Carter}, \bibinfo{person}{Kyle Montague}, \bibinfo{person}{Reem Talhouk}, \bibinfo{person}{Shaun Lawson}, \bibinfo{person}{Hugo Nicolau}, \bibinfo{person}{Ana~Cristina Pires}, \bibinfo{person}{Markus Rohde}, \bibinfo{person}{Alessio~Del Bue}, {and} \bibinfo{person}{Tiffany Knearem}.} \bibinfo{year}{2024}\natexlab{b}.
\newblock \showarticletitle{DCitizens Roles Unveiled: SIG Navigating Identities in Digital Civics and the Spectrum of Societal Impact}. In \bibinfo{booktitle}{\emph{Proceedings of the SIGCHI Conference on Human Factors in Computing Systems}} \emph{(\bibinfo{series}{CHI '24})}. \bibinfo{publisher}{Association for Computing Machinery}, \bibinfo{address}{New York, NY, USA}.
\newblock
\urldef\tempurl%
\url{https://doi.org/10.1145/3613905.3643981}
\showDOI{\tempurl}


\bibitem[Dahl et~al\mbox{.}(2022)]%
        {dahl2022towards}
\bibfield{author}{\bibinfo{person}{Yngve Dahl}, \bibinfo{person}{Kshitij Sharma}, {and} \bibinfo{person}{Dag Svanes}.} \bibinfo{year}{2022}\natexlab{}.
\newblock \showarticletitle{Towards Participatory Design Facilitation as Reflective Practice}. In \bibinfo{booktitle}{\emph{Adjunct Proceedings of the 2022 Nordic Human-Computer Interaction Conference}}. \bibinfo{pages}{1--3}.
\newblock


\bibitem[DCitizens(2022)]%
        {dcitizensproj22}
\bibfield{author}{\bibinfo{person}{DCitizens}.} \bibinfo{year}{2022}\natexlab{}.
\newblock \bibinfo{title}{Fostering Digital Civics Research and Innovation in Lisbon}.
\newblock
\newblock
\urldef\tempurl%
\url{https://doi.org/10.3030/101079116}
\showDOI{\tempurl}


\bibitem[Dow et~al\mbox{.}(2018)]%
        {dow2018between}
\bibfield{author}{\bibinfo{person}{Andy Dow}, \bibinfo{person}{Rob Comber}, {and} \bibinfo{person}{John Vines}.} \bibinfo{year}{2018}\natexlab{}.
\newblock \showarticletitle{Between grassroots and the hierarchy: Lessons learned from the design of a public services directory}. In \bibinfo{booktitle}{\emph{Proceedings of the 2018 CHI Conference on Human Factors in Computing Systems}}. \bibinfo{pages}{1--13}.
\newblock


\bibitem[Hamm et~al\mbox{.}(2024)]%
        {hamm2024does}
\bibfield{author}{\bibinfo{person}{Andrea Hamm}, \bibinfo{person}{Yuya Shibuya}, \bibinfo{person}{Teresa~Cerratto Pargman}, \bibinfo{person}{Roy Bendor}, \bibinfo{person}{Christoph Raetzsch}, \bibinfo{person}{Mennatullah Hendawy}, \bibinfo{person}{Rainer Rehak}, \bibinfo{person}{Gwen Klerks}, \bibinfo{person}{Ben Schouten}, {and} \bibinfo{person}{Nicolai~Brodersen Hansen}.} \bibinfo{year}{2024}\natexlab{}.
\newblock \showarticletitle{What Does' Failure'Mean in Civic Tech? We Need Continued Conversations About Discontinuation}.
\newblock \bibinfo{journal}{\emph{Interactions}} \bibinfo{volume}{31}, \bibinfo{number}{2} (\bibinfo{year}{2024}), \bibinfo{pages}{34--38}.
\newblock


\bibitem[Henriques et~al\mbox{.}(2024)]%
        {Henriques2024}
\bibfield{author}{\bibinfo{person}{Ana~O. Henriques}, \bibinfo{person}{Hugo Nicolau}, \bibinfo{person}{Anna R.~L. Carter}, \bibinfo{person}{Kyle Montague}, \bibinfo{person}{Reem Talhouk}, \bibinfo{person}{Angelika Strohmayer}, \bibinfo{person}{Sarah Rüller}, \bibinfo{person}{Cayley MacArthur}, \bibinfo{person}{Shaowen Bardzell}, \bibinfo{person}{Colin Gray}, {and} \bibinfo{person}{Eleonore Fournier-Tombs}.} \bibinfo{year}{2024}\natexlab{}.
\newblock \showarticletitle{Fostering Feminist Community-Led Ethics: Building Tools and Connections}.
\newblock \bibinfo{journal}{\emph{Proceedings of the 2024 ACM Designing Interactive Systems Conference}} (\bibinfo{year}{2024}).
\newblock
\urldef\tempurl%
\url{https://doi.org/10.1145/3656156.3658385}
\showDOI{\tempurl}


\bibitem[Mandarano et~al\mbox{.}(2010)]%
        {mandarano2010building}
\bibfield{author}{\bibinfo{person}{Lynn Mandarano}, \bibinfo{person}{Mahbubur Meenar}, {and} \bibinfo{person}{Christopher Steins}.} \bibinfo{year}{2010}\natexlab{}.
\newblock \showarticletitle{Building social capital in the digital age of civic engagement}.
\newblock \bibinfo{journal}{\emph{Journal of planning literature}} \bibinfo{volume}{25}, \bibinfo{number}{2} (\bibinfo{year}{2010}), \bibinfo{pages}{123--135}.
\newblock


\bibitem[Menendez-Blanco and Bj{\o}rn(2022)]%
        {menendez2022designing}
\bibfield{author}{\bibinfo{person}{Maria Menendez-Blanco} {and} \bibinfo{person}{Pernille Bj{\o}rn}.} \bibinfo{year}{2022}\natexlab{}.
\newblock \showarticletitle{Designing digital participatory budgeting platforms: urban biking activism in Madrid}.
\newblock \bibinfo{journal}{\emph{Computer Supported Cooperative Work (CSCW)}} \bibinfo{volume}{31}, \bibinfo{number}{4} (\bibinfo{year}{2022}), \bibinfo{pages}{567--601}.
\newblock


\bibitem[Mirra and Garcia(2017)]%
        {mirra2017civic}
\bibfield{author}{\bibinfo{person}{Nicole Mirra} {and} \bibinfo{person}{Antero Garcia}.} \bibinfo{year}{2017}\natexlab{}.
\newblock \showarticletitle{Civic participation reimagined: Youth interrogation and innovation in the multimodal public sphere}.
\newblock \bibinfo{journal}{\emph{Review of Research in Education}} \bibinfo{volume}{41}, \bibinfo{number}{1} (\bibinfo{year}{2017}), \bibinfo{pages}{136--158}.
\newblock


\bibitem[Nouwens et~al\mbox{.}(2022)]%
        {nouwens2022time}
\bibfield{author}{\bibinfo{person}{Minke Nouwens}, \bibinfo{person}{Peter Dalsgaard}, {and} \bibinfo{person}{Amos~Bokonen Blanton}.} \bibinfo{year}{2022}\natexlab{}.
\newblock \showarticletitle{Time and its Study in Design Ideation Processes}. In \bibinfo{booktitle}{\emph{Adjunct Proceedings of the 2022 Nordic Human-Computer Interaction Conference}}. \bibinfo{pages}{1--4}.
\newblock


\bibitem[Olivier and Wright(2015)]%
        {olivier2015digital}
\bibfield{author}{\bibinfo{person}{Patrick Olivier} {and} \bibinfo{person}{Peter Wright}.} \bibinfo{year}{2015}\natexlab{}.
\newblock \showarticletitle{Digital civics: taking a local turn}.
\newblock \bibinfo{journal}{\emph{interactions}} \bibinfo{volume}{22}, \bibinfo{number}{4} (\bibinfo{year}{2015}), \bibinfo{pages}{61--63}.
\newblock


\bibitem[Poplin(2014)]%
        {poplin2014digital}
\bibfield{author}{\bibinfo{person}{Alenka Poplin}.} \bibinfo{year}{2014}\natexlab{}.
\newblock \showarticletitle{Digital serious game for urban planning:“B3—Design your Marketplace!”}.
\newblock \bibinfo{journal}{\emph{Environment and Planning B: Planning and Design}} \bibinfo{volume}{41}, \bibinfo{number}{3} (\bibinfo{year}{2014}), \bibinfo{pages}{493--511}.
\newblock


\bibitem[Reynolds-Cu{\'e}llar and Delgado~Ramos(2020)]%
        {reynolds2020community}
\bibfield{author}{\bibinfo{person}{Pedro Reynolds-Cu{\'e}llar} {and} \bibinfo{person}{Daniela Delgado~Ramos}.} \bibinfo{year}{2020}\natexlab{}.
\newblock \showarticletitle{Community-based technology co-design: insights on participation, and the value of the “co”}. In \bibinfo{booktitle}{\emph{Proceedings of the 16th Participatory Design Conference 2020-Participation (s) Otherwise-Volume 1}}. \bibinfo{pages}{75--84}.
\newblock


\bibitem[Richardson(2023)]%
        {richardson2023failed}
\bibfield{author}{\bibinfo{person}{Dan Richardson}.} \bibinfo{year}{2023}\natexlab{}.
\newblock \showarticletitle{Failed yet successful: OurPlace and Digital Civics}.
\newblock  (\bibinfo{year}{2023}).
\newblock


\end{thebibliography}

\appendix
\end{document}